# Use Me Wisely: AI-Driven Assessment for LLM Prompting Skills Development

**Dimitri Ognibene[1], Gregor Donabauer[2], Emily Theophilou[3], Cansu Koyuturk[1], Mona Yavari[1], Sathya Bursic[1], Alessia Telari[1], Alessia Testa[1], Raffaele Boiano[4], Davide Taibi[5], Davinia Hernandez-Leo[3], Udo Kruschwitz[2] and Martin Ruskov[6]**

[1]Università degli Studi di Milano Bicocca, Italy
[2]University of Regensburg, Germany
[3]Universitat Pompeu Fabra, Spain
[4]Politecnico di Milano, Italy
[5] National Research Council of Italy (CNR), Italy
[6]Università degli Studi di Milano, Italy
*Corresponding author: dimitri.ognibene@unimib.it*

**ABSTRACT:** The use of large language model (LLM)-powered chatbots, such as ChatGPT, has become popular across various domains, supporting a range of tasks and processes. However, due to the intrinsic complexity of LLMs, effective prompting is more challenging than it may seem. This highlights the need for innovative educational and support strategies that are both widely accessible and seamlessly integrated into task workflows. Yet, LLM prompting is highly task- and domain-dependent, limiting the effectiveness of generic approaches. In this study, we explore whether LLM-based methods can facilitate learning assessments by using ad-hoc guidelines and a minimal number of annotated prompt samples. Our framework transforms these guidelines into features that can be identified within learners' prompts. Using these feature descriptions and annotated examples, we create few-shot learning detectors. We then evaluate different configurations of these detectors, testing three state-of-the-art LLMs and ensembles. We run experiments with cross-validation on a sample of original prompts, as well as tests on prompts collected from task-naive learners. Our results show how LLMs perform on feature detection. Notably, GPT-4 demonstrates strong performance on most features, while closely related models, such as GPT-3 and GPT-3.5 Turbo (Instruct), show inconsistent behaviors in feature classification. These differences highlight the need for further research into how design choices impact feature selection and prompt detection. Our findings contribute to the fields of generative AI literacy and computer-supported learning assessment, offering valuable insights for both researchers and practitioners.

## 1. INTRODUCTION

The advent of large language models (LLMs) has empowered the wide-ranging application of Artificial Intelligence (AI) across diverse tasks, all without the need for model retraining or even local deployment, primarily employing prompt-based approaches, which appear to not require a deep understanding of the underlying system, e.g. (Dang et al., 2022). This unprecedented accessibility has sparked the interest of the general public and the media for these models with several attempts to integrate its usage in the most disparate tasks and processes, e.g. (Leinonen et al., 2023). However, significant challenges remain that necessitate education and guidance (Bobula, 2024). Interpreting LLM outputs and discerning their reliability are particularly pressing issues (Chen et al., 2023; Zhou et al., 2024). Additionally, a significant challenge, which we explore in this paper, is the ability to develop effective prompts for novel tasks (Zamfirescu-Pereira et al., 2023).

Users frequently struggle to predict and control the outputs of LLMs for new tasks and domains, hindering their ability to obtain the desired results (Zamfirescu-Pereira et al., 2023). This has recently led to the appearance of some prompting strategies in the literature, e.g. (Yao et al. 2023; Wei et al., 2022). Ideally, these strategies should not only be effective but also easy to adapt to new contexts and straightforward to learn, ensuring that the approach remains accessible and applicable. However, different domains and tasks may need specific considerations and strategies (Sivarajkumar et al., 2024). The preparation of guidelines for the effective usage of LLMs in specific domains (e.g. insurance sector) and tasks (e.g. acceptances, information acquisition, evaluation, payments, internal management tasks) may require the collaboration of a LLM usage expert and a subject domain expert (Raj J et al., 2024). To ensure learners are equipped with the necessary skills, tailored prompting guidelines and examples can be developed, and assessors can monitor their adherence as illustrated in Figure 1. This is similar to what is needed in the development of various socio-technical systems (Mumford, 2000; Fischer et al., 2009), e.g. in an enterprise setting to support effective search where domain experts and technical experts need to work hand in hand (Kruschwitz and Hull, 2017). We define a LLM usage expert to be a practitioner with a deep understanding of LLM technologies, architectures, and nuances. The necessity for such a role emerges from research in prompting (White et al., 2023; Chubb, 2023; Ruskov, 2023). The required effort for the adoption of LLMs on new tasks may not be neglectable compared to that of using a new software tool or programming library (Hernández-Leo et al., 2019; Shrestha et al., 2020). A relevant difference is that often teaching material, handbooks, and exercises for those can be extracted from the development documentation or other structured descriptions of the software system behavior. Instead, the LLM's functioning is not deterministic or easy to interpret (Astekin et al., 2024).

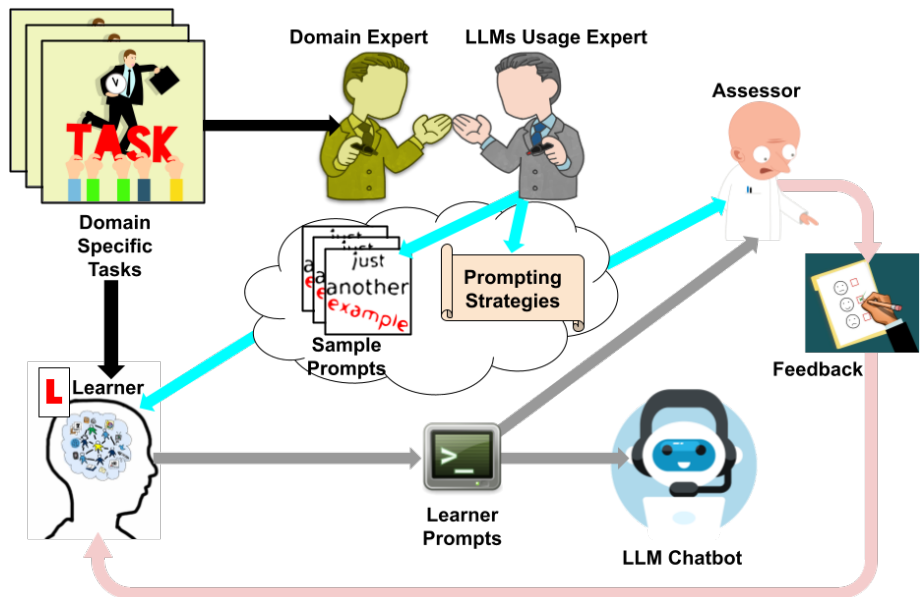

*Figure 1. To effectively tailor prompting guidelines for diverse learners and tasks, domain experts and LLM usage experts should collaborate to develop customized examples, which assessors can then use to monitor and evaluate learners' adherence to these guidelines.*

In recognition of these challenges, integrating automated assessment and monitoring of learners' task-specific prompting skills have the potential to enhance the existing prompting guidelines and learner practice. This approach not only reduces the need of both domain experts and LLMs-usage specialists but also facilitates a seamless transition from training to proficient usage. However, considering the diverse and ever-changing tasks where prompting is employed, automatic assessment, i.e. evaluating and monitoring if learners are improving their prompting skills and adopting the suggested strategies, presents a multifaceted challenge.

This complexity arises from the variability, lack of transparency, and richness of learner-LLM interaction. Traditionally, this would demand a substantial volume of training data to train a domain-specific classification model. However, in this context, with a strong dependence on the tasks and domain, only a limited number of ad hoc created prompting samples will be available. This paper introduces a framework that aims to mitigate the scarcity of training data by leveraging the semantic knowledge encoded within LLMs and swiftly integrating information from both prompting samples and strategies (see Figure 2). It then goes on to examine the effectiveness of the proposed framework, applying it to three distinct sets of prompting guidelines. As a consequence in the context of this framework, we define the following research question to address:

**RQ**: How good are LLMs in assessing the presence of recommended prompting features for ChatGPT prompts?

The initial design choice of the framework is to translate the assessment of prompting strategies into machine learning classification tasks. In this particular case, three diverse prompting strategies were translated into three sets of prompting features, enabling the automatic detection of the adoption of prompting strategies in the learners' prompts. In our work, we define a "feature" as a characteristic of a written prompt that may or may not be present within the prompt's text. For example, if a prompt consists of long and complex sentences, a feature called "complex sentence structure" would be considered present. These features are then used to annotate the sample prompts provided by the experts. Note that the annotation task can be simplified, as these sample prompts are presumably intended to exemplify prompting strategies. The second design choice consists of transforming these features into Assessor Prompt Templates. The framework automatically merges these with labeled sample prompts and learner-generated prompts, creating Feature Detection Prompts. These prompts are then submitted to the LLM to assess whether the specified learner's prompt exhibits the particular feature in question. We experiment with diverse configurations of few-shot LLM detectors, exploring variations in Assessor Prompt Templates, detector settings, ensemble techniques, and the underlying LLMs. Learners' data are collected through online crowdsourcing during a learning activity comprising direct interaction with the LLM and two conditions corresponding to the presentation of two different prompting strategies. Tests with naive learners show the potential of the presented approach.

We summarize our contributions as follows:

- At first, we identify a comprehensive set of features that distinguish between effective and ineffective prompts sent to a LLM
- Next, we formulate assessor prompt templates designed for the development of few-shot detectors. These detectors can be used to classify the presence of the previously identified features in new prompts
- Finally, we conduct an online crowdsourcing experiment to gather prompts generated by learners during a learning activity. We leverage the pre-established assessor prompt templates to assess the performance of LLMs in detecting the specified features during the learning process

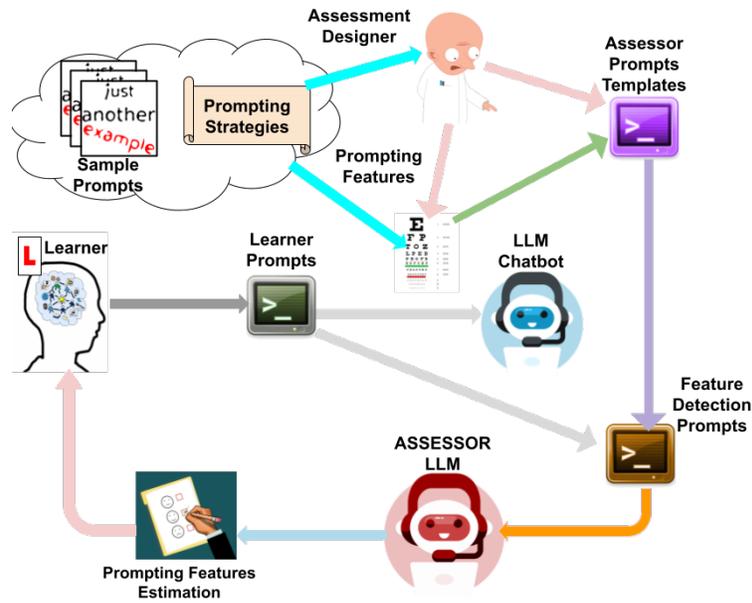

*Figure 2. LLM-based framework for assessing and developing task-specific prompting skills using few-shot learning. This framework enables learners to interact with an LLM, where their prompts are evaluated based on key prompting features. Assessment designers extract and define these features, which are then used by the LLM to assess learner prompts. Few-shot learning is employed to offer feedback and improve learners' prompting abilities, even with limited sample prompts.*

## 2. RELATED WORK

The emergence of generative AI, exemplified by LLM powered chatbots such as ChatGPT, has revolutionized the landscape of educational technology. These tools have paved the way for innovative approaches in providing students with novel learning experiences (Moon et al., 2023; Su et al., 2023). Given the increasing prominence of generative AI, educational institutions are prioritizing its integration into learning materials (Ali et al., 2021; Chen et al, 2023; Kadaruddin, 2023). To maximize the benefits of AI it is essential for schools to introduce comprehensive AI literacy programs that not only cover technical skills but also equip students with a deep understanding of these systems functionality to address prevalent misconceptions (Ottenbreit-Leftwich et al, 2023). In this context, recent research introduces the concept of generative AI literacy as a structured learning path that equips individuals with skills and knowledge to effectively engage with generative AI tools (Annapureddy et al., 2024). To foster generative AI literacy, understanding the core principles and functionalities of generative AI tools is essential. This includes comprehending the underlying concepts, functionalities, and ethical implications to ensure informed and responsible usage. Moreover, the ability to develop effective writing prompts and prompting strategies, known as prompt engineering, has emerged as a crucial competency for harnessing generative AI (Lee et al., 2023). Initial prompting interactions with LLM powered chatbots might seem straightforward. However, mastering effective prompting techniques requires an understanding of identifying common errors in LLM responses. This includes identifying common errors, such as hallucinations, biases, mistakes and incoherence (Fang et al., 2024; Gandolfi, 2024). By recognizing these potential pitfalls, users can devise strategies to address them and systematically evaluate the effectiveness of their approaches (Zamfirescu-Pereira et al., 2023). A first interaction with a LLM powered chatbot often manifests as an emotional rollercoaster, alternating between moments of excitement and frustration (Jiang et al., 2022; Gabbiadini, 2024). Yet, through iterative engagement, users gradually develop a more balanced perspective, gaining a deeper understanding of the LLM's strengths and limitations (Jiang et al., 2022). While a basic understanding of prompting can empower students to critically evaluate the capabilities and limitations of these systems (Theophilou et al, 2023; Denny et al., 2023a), the ability to craft effective prompts has become an essential skill for students to successfully interact with LLM powered chatbots (Denny et al., 2023b). However, this proficiency is often domain or task specific (Sivarajkumar et al., 2024) and usually it is developed through workshops or tutor-led interventions, requiring the guidance or intervention of experienced educators (Denny et al., 2023a; Theophilou et al., 2023; Theophilou et al., 2024).

It is rather evident that there is a learning curve when engaging with LLM powered chatbots, particularly for non-AI literate users who often approach these interactions with expectations rooted in human-human communication (Zamfirescu-Pereira et al., 2023). Learners tend to initiate their interactions with a conversational tone, often extending a greeting to the chatbot (Denny et al., 2023a). Moreover, they occasionally exhibit politeness by incorporating phrases like "please" after issuing commands to the chatbot (Zamfirescu-Pereira et al., 2023). While this inclination is natural, it can sometimes hinder effective communication. For instance, research indicates that both impolite and overly polite prompts can negatively impact LLM performance; impolite language may lead to poor outcomes, while excessive politeness does not guarantee better results (Yin et al., 2023). Additionally, the effectiveness of politeness in prompts is influenced by linguistic and cultural contexts, as models are trained on specific languages and may be sensitive to politeness norms within those cultures.

Therefore, adapting one's conversational style to align with the nature of LLMs becomes crucial for successful interactions (Ostrand & Berger, 2024). Another common issue in interactions with LLM powered chatbots is the prevalence of incomplete prompts, often limited to vague queries. These incomplete prompts frequently lead to further iterative clarifications and extended back-and-forths (Denny et al., 2023a), which can be time-consuming and frustrating for the participant. Additionally, participants may send successive prompts with the intent of enhancing the generated answer, rather than altering the initial query (Denny et al., 2023a).

Teaching students effective prompting strategies can help mitigate the identified challenges. A pilot educational intervention with high school students showed that teaching high-level AI concepts and practical prompting techniques encouraged more structured prompts, leading to a more positive perception of LLM chatbots as clear, natural, and enjoyable to use (Theophilou et al., 2023). Interestingly, this study used ChatGPT to highlight the limitations of LLMs by first allowing students to interact with the tool and attempt to generate specific outputs without any specific prompting knowledge. When their initial prompts proved ineffective, it underscored the challenges of using LLMs without proper guidance. Researchers then demonstrated effective prompting techniques, which enabled students to craft more successful prompts, leading to improved outputs and a reduction in negative sentiments towards AI.

While initial attitudes toward AI can vary widely (Sánchez-Reina et al., 2024), a positive experience with LLMs can positively influence participants' perceptions and potentially increase their willingness to utilize AI solutions more frequently (Kwak et al., 2022; Theophilou et al., 2023). This underscores the significance of providing users with appropriate guidance and education to fully harness the capabilities of AI systems (Zamfirescu-Pereira et al., 2023). Recognising this, documentation and practical strategies could be offered to users to provide them with information on how to design effective prompts. This would help them understand the potential pitfalls and improve their ability to create successful prompts (Jiang et al., 2022).

In an effort to empower novice AI users to effectively utilize LLM powered chatbots, numerous prompting guides, patterns, guidelines, and tools have emerged in the literature (Giray, 2023; Akin, 2023; Denny et al., 2023a; White et al., 2023). The common goal of these resources is to provide users with actionable strategies for crafting effective prompts that elicit desired responses from LLMs. In addition to providing users with guidelines on prompting techniques, further educational practices can be employed, such as feedback techniques, to offer users constructive feedback on their proficiency, a critical aspect of effective teaching and learning.

Feedback plays a pivotal role in shaping performance and fostering improvement, providing learners with valuable insights into their performance, highlighting both task accomplishment and areas for enhancement (Vollmeyer & Rheinberg, 2005). Feedback is typically provided following the assessment of students' work. This process is often initialized with an assessor that puts together assessment guides or rubric scales to facilitate the evaluation process, allowing evaluators to assess learning outcomes based on pre-established criteria (Reddy & Andrade, 2010; Dawson, 2017). These tools provide a structured framework for evaluating student performance, ensuring consistency and objectivity throughout the assessment process.

Automated assessment powered by AI tools is an emerging approach (del Gobbo et al., 2023); however, as with any human evaluator, it necessitates the development of a set of predefined templates, features, and guidelines formulated by an assessor to guide AI towards accurate assessment of student work (Dawson, 2017). In the context of prompting education, the evaluation of users' prompts would require the tracking of a set of learning curves to measure the learning progress of users. On this line, White et al. (2023) present a comprehensive catalog of prompting patterns and their combinations to enhance LLM performance. Their categorization encompasses five key areas: input semantics, output customization, error identification, prompt improvements, and interaction, aiming to address various aspects of prompt engineering. This categorization holds the potential to serve as a foundation for assessing prompting proficiency, as it reflects a high-level standard of effective prompt engineering.

While these guidelines have yet to be evaluated in an educational setting, they have the potential to be adapted to specific tasks by proficient LLM users and domain experts as well as guiding novice AI users towards learning to write improved prompts and receive their desired outcome. By studying the patterns of successful prompt design in educational contexts, educators and learners can gain valuable insights into prompt engineering proficiency and identify areas for improvement. Automation of this assessment could be facilitated by the LLM itself (Nguyen et al., 2023), an innovative approach that has yet to be explored in the existing literature. Controlled environments, similar to the study by Koyuturk et al. (2023), can be used to isolate the prompt design process and systematically assess the impact of different prompt strategies on learning outcomes.

Drawing upon the insights from the existing literature and recommendations for effective prompting strategies, we propose a set of features for evaluating prompting techniques within the LLM context of ChatGPT, and evaluate its feasibility for assessing users' prompting skills. Specifically, we conduct a comparative analysis of various feature sets. These sets include the patterns proposed by White et al. (2023) and others crafted ad hoc to aid in supporting users in utilizing generative AI chatbots. Our overall objective is to monitor and identify enhancements in their prompting skills as they progress in learning to prompt. To achieve this, in this paper we apply the proposed automatic assessment methodology to a diverse set of features and different versions of ChatGPT (to get an idea as to how performance differs between different capable models). This will help us to evaluate the effectiveness of our methodology and identify any potential limitations.

## 3. METHODOLOGY

### 3.1. PARTICIPANT RECRUITEMENT

To assess the feasibility of employing an LLM to evaluate the existence of specific features within prompts, the initial step of this study was designed to facilitate the collection of prompts through the Prolific platform (Palan and Schitter, 2018). This resulted in the recruitment of 63 participants (representing a higher number than in related studies, e.g. Denny et al. (2023a), Zamfirescu-Pereira et al. (2023), and Theophilou et al. (2023)) who consented to participate in an informal learning activity. Participants were informed about the research objective of the study and had the choice to withhold their data from being processed.

All recruited participants were located in the UK or the US (m=26, f=37). Their age ranges from 18 to 30 years (mean=25.3, sdev=3.18). Educational backgrounds varied, from secondary school to doctorate or postgraduate degree, with the majority having completed high school (33.3%) and Bachelor's Degree (44.4%). The average self-reported knowledge of digital technology, measured on a 7-point Likert scale (ranging from 1-not familiar to 7-very familiar) was 4.35. Regarding prior experience with ChatGPT, all participants indicated that they have either 'never' or 'once or twice' used ChatGPT.

## 3.2. LEARNING ACTIVITY AND STUDY PROCEDURE

On the Prolific platform, participants were instructed to engage in a learning activity aimed at mastering prompting strategies. To achieve this, we offered participants a comprehensive catalog of prompting patterns and their combinations, inspired by the prompt categorization developed by White et al. (2023). This approach was chosen to equip participants with effective techniques that would enhance their interactions with LLM powered chatbots and maximize their performance. We adopt the study design presented in one of our previous studies (Theophilou et al., 2023) which structured the educational intervention into distinct sections. This approach not only facilitated the collection of prompting techniques but also provided participants with comprehensive education on various prompting strategies. However, we adapted the study design to accommodate an online format as opposed to the described school-based setting. As a result, a learning activity consisting of different stages was formulated as shown in Figure 3.

In the first stage, participants responded to questions related to their experience with generative AI tools, to help us determine their level of proficiency with prompting techniques. During the data analysis, we selected data coming from participants that did not have experience with generative AI. This was done to select participants who would potentially benefit the most from prompt strategy learning activities. However, for a comprehensive understanding of how such learning activities can influence prompting skills across all levels of experience, further experiments are necessary as also outlined in more detail in the limitations section of this paper. The second phase introduced participants to a video covering key concepts related to AI applications, LLMs, and the distinction between human intelligence and artificial intelligence. To verify that participants had watched the entire video, a question was included within the page, with its answer concealed within the video's content.

In the next phase, participants received a set of instructions to complete a prompting activity within an ad hoc chatbot interface that was powered by the ChatGPT API. In particular, participants were tasked to instruct the chatbot to act as a personal teacher and to provide them with information about three specific social media topics of their choice. Moreover, the instructions included requirements such as: "explain in a friendly interactive way the potential threats and related mechanisms of social media", "avoid long bulleted lists", "offer explanations and support and provide references". The ultimate goal of the task was to foster a more engaging interaction where the chatbot provided responses in a conversational manner, rather than merely delivering information in a structured, encyclopedia-like format. Within the chat interface, participants received a code to provide in the questionnaire to ensure the completion of the task. The prompting activity was repeated two more times with the goal of improving the interaction in each iteration. After the third iteration, participants were instructed to watch a video providing information in regard to prompting strategies. Then, participants completed the chatbot interaction section three more times. Overall, each participant submitted six initial prompts to the Chatbot, resulting in 378 samples.

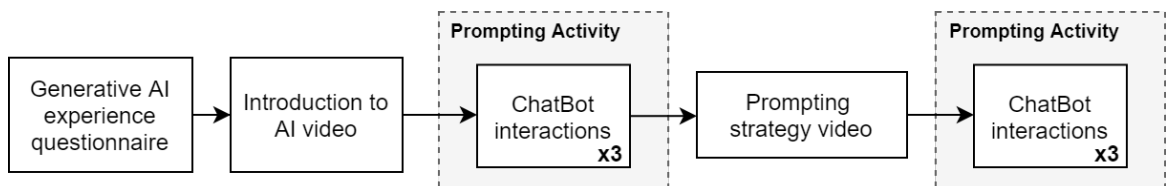

*Figure 3. An overview of the procedure of the study.*

### 3.3. DATA COLLECTION AND ANALYSIS

### 3.3.1. CHATBOT INTERFACE

To allow the collection of participants' prompts during the learning activity, we adopted an approach from our previous work (Koyuturk et al., 2023), which involved designing an interface based on the ChatGPT API. This decision was driven by the need for an ad-hoc platform that could seamlessly capture interactions. This web application is built with TypeScript, utilizing React, NextJS, and SQLite. Upon accessing the interface, users are presented with an initial task description and a start button. After clicking the "start" button, they are redirected to a chat interface, allowing them to engage with the ChatGPT API. An example for a conversation within the interface is shown in Figure 4. Each learner is uniquely identified by a participant ID, and all messages exchanged are logged in a SQL database for later analysis. In the background, the database model saves the users' interaction history during the experiment as a chat object where each one of them consists of multiple message objects. The main part of a message object is the textual input of the user. The responses of ChatGPT are also saved in the database and used as the interaction history for future prompts of the same user. Apart from that, the backend saves meta information, like timestamps, that are related to the experiment and that are associated with a chat.

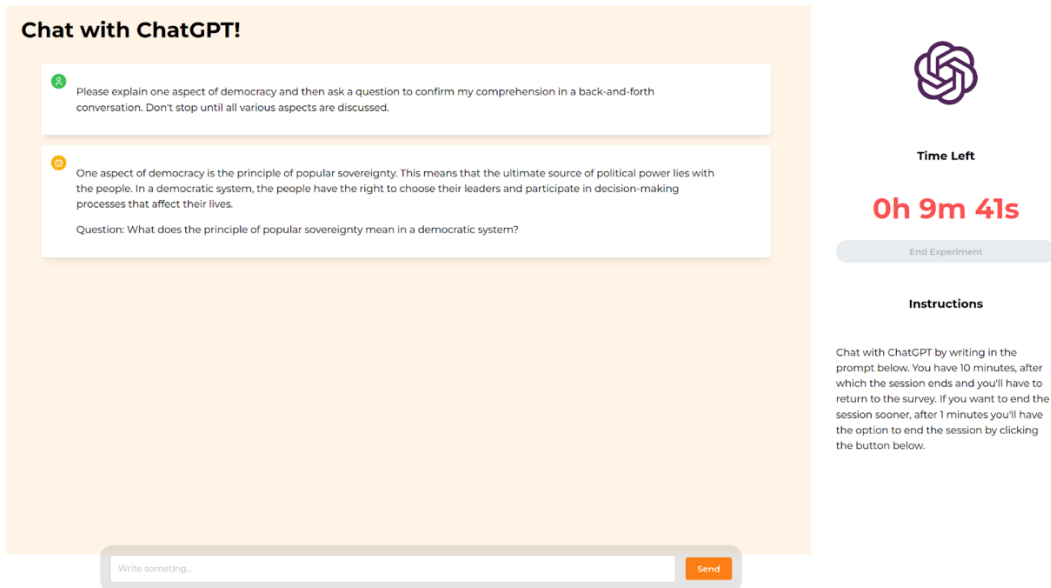

*Figure 4. Screenshot of the interface with an ongoing conversation.*

### 3.3.2. FEATURE SELECTION AND ANNOTATION

To define features that characterize effective prompts, we perform a feature selection and annotation process that consists of three sequential steps: (1) we start by investigating three sets of potential features, drawing inspiration from the existing literature (White et al., 2023) and suggestions by ChatGPT itself (baseline); (2) Subsequently, we annotate sample prompts with the identified feature categories. This annotation phase helps to associate prompts with relevant features, allowing for an informed selection; (3) We finally select the most relevant set of features, derived from the annotations. This step refines the candidate features, focusing on those that are most relevant to the context.

The candidate feature sets emerge from three distinct approaches for categorizing features conducive or detrimental to LLM prompts. The first method involves an in-depth exploration of the feature space, yielding 21 features. The second method adopts the prompting patterns proposed by White et al. (2023), encompassing 14 features. The third approach involves asking ChatGPT directly for insights, resulting in the identification of 7 distinct features. In summary, our candidate feature space comprises 42 distinctive features. To assess the distribution of these features within same-domain prompts, we asked three annotators to annotate a set of 20 high-quality sample prompts derived from a previous experiment (Theophilou et al., 2023) with the features identified in step one. These samples will later be used as the training set for selecting few-shot examples. Additionally, we sampled a subset of 40 prompts from the prompting experiments conducted with the 63 participants, selected based on the previously mentioned criteria (such as prior experience with ChatGPT). These prompts are also annotated with the same set of features and will subsequently serve as a test set.

To avoid overlapping feature spaces and eliminate those that might only have marginal impact, we performed a feature reduction operation. This process involved calculating the odd ratio value, a measure of the discriminative capacity of each feature, and ensuring high agreement among annotators. Consequently, a final set of 17 distinct features was selected. To measure the inter-annotator agreement for the annotation of all sample prompts using the final feature set, we employed Fleiss Kappa, yielding an exceptionally high value of $\kappa = 0.992$, indicative of strong consensus among annotators. The final set of the 17 features we identified by applying this methodology is listed below in Table 1.

| Feature Name | Feature Description |
|---|---|
| Topic - concise | cursorily description of the topic with few details |
| Topic - broken down | broken down description of the topic with details |
| 1 Goal | description of exactly one countable distinct goal |
| 2 Goals | description of exactly two countable distinct goals |
| >2 Goals | description of more than two countable distinct goals |
| AI role play | assigning a role to the language model and how it should be played |
| Role form/context | additional contextual information about the role of the language model, the user, or the environment |
| Meta Process-related | description of process information, e.g. repeat procedure, continue, etc. |
| Simple sentence structure | usage of simple sentences, even if many |
| Complex sentence structure | complex sentence structure |
| Act As Persona - Persona Pattern | instructing the language model to act as a specific persona |
| Provide Outputs - Persona Pattern | instructing the language model about the expected output by using keywords like: teach, explain, etc. |
| Pattern Order - Persona Pattern | instructions about language model behavior before instructions about expected output |
| Strict Separation Role Vs Output - Persona Pattern | instructions about language model behavior before instructions about expected output and they are separate statements divided by punctuation or conjunctions |
| Ask Me Questions - Flipped Pattern | instructing the language model to ask back questions |
| Condition Stop - Flipped Pattern | including a stop condition for the conversation with the language model until a condition is met or a goal is achieved |
| Form-Flipped Pattern | instructing the language model to answer in a specific form, e.g. ask me questions one at a time, two at a time, etc. |

*Table 1. Features and short description.*

### 3.3.3. FEW-SHOT LEARNING

We employ a few-shot learning approach using LLMs to detect the features in new prompts given during the learning activity. In particular, we use GPT-3.5 Turbo, GPT-3.5 Turbo Instruct, GPT-4, and an ensemble approach (which has proven to be effective in other NLP classification tasks, such as hate speech detection (Zimmerman et al., 2018)) outlined in the following section. We assess various models because the level of difficulty in identifying the proposed features is yet unclear, thereby having an influence on the required capability of the model. This analysis allows us to gain insights into the diverse capabilities exhibited by different GPT versions in tackling the task.

The different models we use are all from the GPT family: text-davinci-003 is an early GPT-3 version which can be used for single-turn text generation but is slower than the newer GPT-3.5 Turbo model. This model instead is optimized for cost-efficiency and speed, making it a good fit for conversational applications and general-purpose tasks but with limitations in handling complex reasoning. GPT-3.5 Turbo Instruct is a variant of GPT-3.5 Turbo, optimized for following explicit instructions rather than dialogue, making it suitable for task-specific outputs as it also is the case in our setup. Finally, GPT-4 is the most advanced model we use. It has shown the best performance on different benchmarks and can be used for reasoning or to handle complex, multi-step tasks with more focus on context. The differences between the models we use can help us to identify which types of models are most powerful for few-shot learning on the given task, for example whether it is helpful to use a model that was fine-tuned on instruction following or whether similar results can be achieved with different generations of models. While new LLMs are continuously released, we decided to focus on well-established versions with proven capabilities. Future work could also explore evaluating other models on the task, including GPT-4o or open models such as Llama (Touvron et al., 2023) and Mistral (Jiang et al., 2023).

Few-shot learning is a learning approach where the model is given at inference a small number of demonstrations of each new task it is asked to perform (Brown et al., 2020). Note that the weights of the model are not updated, thus, the model must use its prior knowledge to generalize from these examples to perform the task. As Brown et al. (2020) have shown, large language models excel at zero-shot, one-shot, and few-shot learning tasks, frequently matching performances of fine-tuned models. However, it has been shown that this type of few-shot learning can be unstable (Zhao et al., 2021; Ye & Durrett, 2022). The choice of prompt format, training examples, their number, or even their order all influence the performance and expose biases inherent in the model (Webson and Pavlick, 2021). Nonetheless, few-shot learning is being explored due to its speed, low cost and data efficiency in solving custom tasks (Ahmed et al., 2022; Wei et al., 2022). In particular, we could create detectors for all 17 identified features creating just one Assessor Prompt Template and automatically inserting in it the feature description. As LLMs improve, their ability for few-shot learning is also expected to increase, where GPT-4 gives better performance than GPT3.5 and text-davinci-003. In our case, for each feature and a prompt to evaluate, the model is given:

- a description of the feature;
- a number of positive examples of the feature, i.e. prompts where the feature is present;
- a number of negative examples of the feature, i.e. prompts where the feature is not present;
- the prompt to evaluate

The model is explicitly or implicitly given the task of classifying the presence of the said feature in the prompt to evaluate by directing it to give a yes/no answer as output. To arrive at a classification score, we take the following approaches:

- Direct classification: here, the model's textual output is parsed and converted into a boolean classification. This is enabled by GPT-4 and GPT-3.5 Turbo API.
- Probabilistic classification: the model's log probability for the first token is recorded. This allows for assessing the model's confidence in its output. This is enabled by GPT-3.5 Turbo Instruct API.

- Ensemble classification: outputs of up to 19 different runs are combined using a majority approach when direct classification is used or also mean and max log probability when probabilistic classification is used

An example prompt for the probabilistic classification approach can be seen in Figure 5.

| | |
|---|---|
| **Prompting Feature Description** | *"Me: Answer with Yes or No if this feature:*<br>*additional contextual information …*<br><br>*is present in the following prompt:* |
| **Negative Prompt Example** | *Explain the negative sides of social media use without using bulletins …*<br>**You: No** |
| **Positive Prompt Example** | *Me: and in the following prompt?*<br>*I'm a student!  Could you be my super-cool "teacher" for a bit and …*<br>**You: Yes** |
| **Learner Prompt** | *Me: and in the following prompt?*<br>*Hello! Please try to act like my teacher teaching me disadvantages of social media …*<br>**You:"** |

*Figure 5. Example prompt (right side) for probabilistic classification.*

# 4. RESULTS

In response to our research question, Table 2 shows the performance of the best systems across the 17 features adopted for the task. Several few-shot prompting strategies were tested on the GPT-3.5 Turbo models but did not produce interesting results even when cross-validating on the train dataset. In our approach, cross-validation on the training set involves selecting a subset of samples from the training set to serve as few-shot examples. These few-shot examples are then employed in a prompt to predict features in a test prompt, which is also derived from the same training set discarding the previously selected samples. Indeed, the mean accuracy of 0.55 we can observe using this approach is very low. The feature specific ensemble approach instead provided improved performance. Its approach of aggregating votes on a feature level resulted in noticeably better metric scores. In our experiments, ensembles with few voting models, such as 9 or 11, showed the best performance. This may be a consequence of a limitation in our settings: the limited size of the dataset. The improved performance with ensemble methods demonstrates how the issue of variability in few-shot learning can be mitigated through these approaches.

The GPT-4 model was the best-performing model when cross-validating on the training set, reaching 0.76 mean accuracy over 3 runs. When testing on the collected test dataset, the performance went down to 0.69. We suppose that this is partly due to the different statistics of the training and test datasets reported in Table 3. We report on additional metrics for these runs, such as macro F1-score, precision and recall, in the Appendix in Tables 5, 6 and 7. This correctly reflects the application condition where the training set may be produced by few more expert users working for a longer time to achieve better results while the testing set may be the result of few practice sessions for students.

|  | Mean (by feature) | Stdev (by feature) | Max (by feature) | Min (by feature) |
|---|---|---|---|---|
| gpt-3.5 logprob best prompt train (mean on 5 runs) | 0.55 | 0.09 | 0.71 | 0.40 |
| gpt-3.5 best prompt train (mean on 5 runs) | 0.55 | 0.20 | 0.88 | 0.17 |
| gpt-3.5 ensembles train (mean on 5 runs) | 0.64 | 0.14 | 0.91 | 0.43 |
| gpt-4 mean train (3 runs) | 0.76 | 0.14 | 0.98 | 0.47 |
| gpt-4 stdev train (3 runs) | 0.045 | 0.06 | 0.18 | 0.01 |
| gpt-4 mean test (3 runs) | 0.69 | 0.31 | 0.90 | 0.40 |
| gpt-4 stdev test (3 runs) | 0.04 | 0.04 | 0.09 | 0.00 |

*Table 2. Performance (accuracy) of the few shot systems in detecting the features in the students' prompts.*

To better understand the system performance, we also analyze feature by feature. We find substantial differences between model performance across them, as shown in Table 4. This was not affected by our experimentation with diverse prompt configurations or changing the number of samples. While this setting may be suitable for automatic prompting optimization, it shows the strong difference between LLM models' biases (Zhou et al., 2022). For additional metrics, including macro F1-score, precision, and recall calculated at the feature level, we refer to Tables 8, 9, and 10 in the Appendix.

Features that were particularly difficult to deal with are related to countables. We had formulated these as classification tasks by separating the different options (e.g. separate classes for single command, two commands, three or more). Possibly, better performance can be obtained by changing the question format asking for a direct count, thus adopting a regression approach.

Only in a few conditions, the numbers reported for GPT4 were lower than those achieved by GPT3-based systems in the training phase. In some cases, this reflected the annotator's uncertainty before the alignment process. In particular, the simple sentence structure had shown low agreement among the annotators leading to a stricter definition of the feature. This indicates that more capable models tend to produce better results on our feature classification task.

|  | Single Command | Two Commands | Three or more Commands | Simple Sentence Structure | Ask Questions | Stop Questions |
|---|---|---|---|---|---|---|
| Training Dataset | 0.10 | 0.15 | 0.75 | 0.70 | 0.85 | 0.30 |
| Test Dataset | 0.35 | 0.20 | 0.45 | 0.90 | 0.35 | 0.00 |

*Table 3. Training and test dataset statistics.*

## 5. DISCUSSION

The advent of LLM powered chatbots, with their capabilities together with their unpredictability and bias, requires the development of new educational tools and methods to enable a wide population to access them. Emerging concepts like generative AI literacy emphasize the importance of acquiring prompting skills, enabling students to use these systems proficiently (Annapureddy et al., 2024; Lee et al., 2023). However, effective LLM usage needs to be strongly adapted to the specific context and task at hand (Sivarajkumar et al., 2024). Thus, automating the monitoring and assessment of learners' interaction cannot rely on the reuse of large corpora and needs context-specific material that can be too expensive to produce at scale for the few learners.

| | Single Command | Two Commands | Three or more Commands | Simple Sentence Structure | Ask Questions | Stop Questions |
|---|---|---|---|---|---|---|
| gpt-3.5 logprob best prompt train (mean on 5 runs) | 0.16 | 0.24 | 0.66 | 0.71 | 0.81 | 0.29 |
| gpt-3.5 best prompt train (mean on 5 runs) | 0.40 | 0.44 | 0.61 | 0.55 | 0.67 | 0.54 |
| gpt-3.5 ensembles train (mean on 5 runs) | 0.43 | 0.49 | 0.70 | 0.71 | 0.88 | 0.54 |
| gpt-4 mean train (3 runs) | 0.47 | 0.51 | 0.83 | 0.57 | 0.91 | 0.98 |
| gpt-4 stdev train (3 runs) | 0.03 | 0.01 | 0.03 | 0.03 | 0.04 | 0.03 |
| gpt-4 mean test (3 runs) | 0.40 | 0.90 | 0.65 | 0.85 | 0.55 | 0.85 |
| gpt-4 stdev test (3 runs) | 0.06 | 0.03 | 0.08 | 0.09 | 0.22 | 0.02 |

*Table 4. Performance (accuracy) of the few shot systems on specific features.*

In particular with our research question, we investigated the possibility of using small-sized training samples in a few-shot learning framework based on LLM to evaluate students' prompting strategies. This is inherently challenging because LLM prompting strategies have been developed also to circumvent LLM limitations in interpreting learner inputs (Nam et al., 2024). Yet an adequate few-shot learning framework combining desired features for learner commands integrated with expert-provided examples and annotations may provide a reasonable solution.

Our experimental results show that only the most recent version of GPT provides sufficient performance to support this approach. The annotated samples from the experts could allow the adoption of an automatic prompting method for those features that appear more problematic with manual exploration of prompt configurations (Zhou et al., 2022; Pryzant et al., 2023). However, this automatic prompting might be computationally, temporally, data, and economically expensive and may not solve some specific issues. In particular, we noticed substantial differences in how different GPT models respond to different questions. It may be the case that different formulations for the assessment of the prompting strategies may result in better performance but can hardly be faced by current automatic prompting frameworks. For example, evaluating the adoption of a strategy through a classification or regression measurement appears to strongly affect model performance and cannot be yet solved with automatic prompting strategies. It is thus important for the implementation of this assessment framework to be aware of these different aspects, which would require the expert or the teacher to reformulate or adapt the description of the prompting strategy to achieve higher accuracy.

Another important aspect is the difference between the training set created by the expert and the one resulting from student practice. While the experts may provide prompt samples reflecting a different level of adoption of the prompting strategies, the differences in the statistical distribution of the features as well as the linguistic formulations may affect the automatic assessment performance. A viable solution could be to manually annotate samples from the learners' prompts dataset to obtain a more aligned training set.

## 6. CONCLUSION AND FUTURE WORK

Facilitating the adoption of LLM-based interactions and improving generative AI literacy are novel and important challenges. Although LLM technology is continuously advancing, the ability to craft efficient and clear prompts is becoming increasingly vital (Knoth et al., 2024). This skill will stay relevant even in the presence of new techniques like automatic prompting because they require creating specific datasets for each task and domain.

On the other hand, improvement in LLM models and newly arising prompting strategies may simplify the prompting task, removing some idiosyncratic behaviors (Schulhoff et al., 2024). Still, as in software engineering or in design requirements elicitation, learning to communicate own requests, especially to artificial systems, is likely to keep its importance or increase its importance with the diffusion of generative AI and its continuous application to new tasks.

While we left out several important aspects of the prompting process, such as interpreting generative AI output and refining the request over multiple exchanges, the focus on correctly formulating the first request is an important use case, for example, to extend the skills of workers in a company with related tasks. The task is harder than it seems and often needs external aid. Indeed, during the data collection, several participants dropped the task very early because they felt they could not improve their prompts after the first two attempts.

As mentioned earlier, applying our study protocol to a more diverse set of users with different levels of prompting experience (e.g. novices versus experienced users) could lead to more comprehensive insights which is why we consider it as an important future research question. This could also include investigating how user background and previous skills are correlated with prompt improvement.

Supporting the adoption of LLMs for a group of related tasks is an interesting path and the approach we proposed here would also help understand the students' attitude. The same approach can be applied to other questions, such as estimating the efficacy of the prompt for the task instead of the feature.

Important insights from our experiments are the strong dependence of the performance on the LLM and feature pairs, as well as the importance of translating the prompting guidelines into features with a format that helps LLM tasks. In the future, we aim to analyze different aspects of the interaction with the LLMs, such as comparing the different learning trajectories of the students and refine and compare the effectiveness of different prompting guidelines. Additionally, our study creates a basis for LLM-based prompt evaluations using a fixed set of well-established models (GPT3, GPT3.5 and GPT4). In the future, experiments could expand to include a broader range of models to determine which are best suited for our task.

Finally, practical implications the implementation of our approach causes are costs due to the API usage of GPT models, which can quickly add up with a high number of students since usage is often charged per token. As an alternative, open models like Llama could be deployed independently, though this would require costly technical infrastructure and expertise. Additionally, integrating these models into established learning platforms can be difficult due to their closed ecosystems. Furthermore, LLMs still have limitations, such as hallucinations and instability during few-shot learning, as previously mentioned. Overall, implementing our approach within existing systems presents several challenges, offering plenty of opportunities for further research.

While this study does not extensively address the ethical implications of generative AI technologies, it is important to recognize that biases in LLM outputs can affect prompt engineering (Fang et al., 2024). Understanding and addressing these ethical concerns can inform the development of more effective and responsible prompting strategies, specifically by designing prompts that mitigate bias in LLM outputs (Yang et al., 2023). For instance, awareness of potential biases could lead to prompts specifically designed to reduce such issues, thereby increasing the reliability and fairness of LLM responses (Kamruzzaman & Kim, 2024).

On a final note, transparency into how generative AI models generate their answers potentially inform the development of better prompts. Users can see how an answer was generated, helping to identify and mitigate biases and use the generated outputs responsibly (Franzoni, 2023). This knowledge can also empower users to assess the model's capabilities and decide when to rely on its results, ultimately leading to more effective and responsible prompt engineering. Future work could benefit from a more explicit integration of ethical considerations into the technical aspects of prompt engineering to ensure the responsible use of generative AI tools.

## 7. LIMITATIONS

In our study, we deliberately included only participants who indicated they have no prior experience with ChatGPT. We chose this group because they are expected to experience the greatest positive impact from prompt learning activities. However, we acknowledge that this decision limits the generalizability of our findings. Users with varying levels of experience, particularly more advanced or frequent users of ChatGPT, could also benefit from such strategies, in potentially different ways. Incorporating experienced users into future studies could provide more insights into how prompt learning techniques affect users at different expertise levels. This would allow us to better understand the nuances in learning curves, and the varying degrees of benefit across user types. Therefore, we consider applying our study protocol to a more diverse set of experienced users as an important next step in our work.

Few-shot learning, which we used in our study for prompt feature detection, provides a simple and cost-efficient way to create classifiers for different tasks. However, we acknowledge that this approach can be unstable which is why such models should for example not be used to replace expert annotations (Soboroff, 2024). While this is a common challenge with LLMs and addressing it is beyond the scope of this work, strategies to mitigate this issue are important, also in the context of our research. As a solution to make the approach more robust we used ensemble methods with majority voting across multiple predictions. Another potential way could involve increasing the number of annotated samples per feature, providing more examples during few-shot learning.

## 8. ETHICAL CONSIDERATIONS

While conducting our investigation about the utilization of LLMs for effective prompt learning assessments, ethical considerations have been an important factor. We were ensuring informed consent of participants and performed data anonymization before running the few-shot feature detection experiments. As part of these experiments, we evaluated different models to address biases in LLM model selection and utilization (Gallegos et al., 2024). However, we recognize limitations in generalizing LLM-based methods across tasks and acknowledge the need for an understanding of task dependency (Ge et al., 2023). We also acknowledge ongoing ethical responsibilities, especially important for future research, and emphasize continuous monitoring of ethical considerations in the evolving landscape of LLMs in education (Yan et al., 2024).

## ACKNOWLEDGEMENTS

This work was supported by the Volkswagen Foundation (COURAGE project, no. 95567) as well as by PID2020-112584RB-C33, PID2023-146692OB-C33, CEX2021-001195-M funded by MICIU/AEI/10.13039/501100011033 and SGR 00930. DHL (Serra Húnter) also acknowledges the support by ICREA Academia.

## APPENDIX

*Table 5. Macro F1-scores for the Experiments presented in Table 2.*

|  | Mean (by feature) | Stdev (by feature) | Max (by feature) | Min (by feature) |
|---|---|---|---|---|
| gpt-3.5 logprob best prompt train (mean on 5 runs) | 0.51 | 0.08 | 0.62 | 0.36 |
| gpt-3.5 best prompt train (mean on 5 runs) | 0.46 | 0.14 | 0.62 | 0.16 |
| gpt-3.5 ensembles train (mean on 5 runs) | 0.53 | 0.11 | 0.74 | 0.39 |
| gpt-4 mean train (3 runs) | 0.73 | 0.15 | 0.98 | 0.42 |
| gpt-4 mean test (3 runs) | 0.66 | 0.18 | 0.88 | 0.38 |

*Table 6. Macro Precision for the Experiments presented in Table 2.*

|  | Mean (by feature) | Stdev (by feature) | Max (by feature) | Min (by feature) |
|---|---|---|---|---|
| gpt-3.5 logprob best prompt train (mean on 5 runs) | 0.57 | 0.06 | 0.65 | 0.44 |
| gpt-3.5 best prompt train (mean on 5 runs) | 0.62 | 0.16 | 0.81 | 0.29 |
| gpt-3.5 ensembles train (mean on 5 runs) | 0.64 | 0.19 | 0.98 | 0.36 |
| gpt-4 mean train (3 runs) | 0.76 | 0.13 | 0.98 | 0.51 |
| gpt-4 mean test (3 runs) | 0.73 | 0.16 | 0.93 | 0.51 |

*Table 7. Macro Recall for the Experiments presented in Table 2.*

|  | Mean (by feature) | Stdev (by feature) | Max (by feature) | Min (by feature) |
|---|---|---|---|---|
| gpt-3.5 logprob best prompt train (mean on 5 runs) | 0.56 | 0.05 | 0.63 | 0.45 |
| gpt-3.5 best prompt train (mean on 5 runs) | 0.57 | 0.07 | 0.67 | 0.41 |
| gpt-3.5 ensembles train (mean on 5 runs) | 0.58 | 0.07 | 0.71 | 0.48 |
| gpt-4 mean train (3 runs) | 0.76 | 0.13 | 0.98 | 0.51 |
| gpt-4 mean test (3 runs) | 0.71 | 0.13 | 0.89 | 0.51 |

*Table 8. Macro F1-scores of the few shot systems on specific features.*

| | Single Command | Two Commands | Three or more Commands | Simple Sentence Structure | Ask Questions | Stop Questions |
|---|---|---|---|---|---|---|
| gpt-3.5 logprob best prompt train (mean on 5 runs) | 0.16 | 0.24 | 0.61 | 0.49 | 0.54 | 0.26 |
| gpt-3.5 best prompt train (mean on 5 runs) | 0.36 | 0.39 | 0.58 | 0.44 | 0.51 | 0.51 |
| gpt-3.5 ensembles train (mean on 5 runs) | 0.39 | 0.44 | 0.67 | 0.42 | 0.70 | 0.48 |
| gpt-4 mean train (3 runs) | 0.46 | 0.42 | 0.83 | 0.54 | 0.85 | 0.98 |
| gpt-4 mean test (3 runs) | 0.38 | 0.88 | 0.64 | 0.81 | 0.49 | 0.85 |

*Table 9. Macro Precision of the few shot systems on specific features.*

| | Single Command | Two Commands | Three or more Commands | Simple Sentence Structure | Ask Questions | Stop Questions |
|---|---|---|---|---|---|---|
| gpt-3.5 logprob best prompt train (mean on 5 runs) | 0.31 | 0.56 | 0.78 | 0.55 | 0.54 | 0.29 |
| gpt-3.5 best prompt train (mean on 5 runs) | 0.48 | 0.54 | 0.61 | 0.44 | 0.53 | 0.53 |
| gpt-3.5 ensembles train (mean on 5 runs) | 0.50 | 0.57 | 0.73 | 0.36 | 0.93 | 0.48 |
| gpt-4 mean train (3 runs) | 0.63 | 0.51 | 0.83 | 0.55 | 0.93 | 0.98 |
| gpt-4 mean test (3 runs) | 0.51 | 0.93 | 0.65 | 0.91 | 0.55 | 0.85 |

*Table 10. Macro Recall of the few shot systems on specific features.*

| | Single Command | Two Commands | Three or more Commands | Simple Sentence Structure | Ask Questions | Stop Questions |
|---|---|---|---|---|---|---|
| gpt-3.5 logprob best prompt train (mean on 5 runs) | 0.41 | 0.58 | 0.64 | 0.53 | 0.55 | 0.43 |
| gpt-3.5 best prompt train (mean on 5 runs) | 0.46 | 0.60 | 0.59 | 0.45 | 0.55 | 0.54 |
| gpt-3.5 ensembles train (mean on 5 runs) | 0.52 | 0.71 | 0.68 | 0.51 | 0.65 | 0.48 |
| gpt-4 mean train (3 runs) | 0.70 | 0.51 | 0.82 | 0.56 | 0.79 | 0.98 |
| gpt-4 mean test (3 runs) | 0.51 | 0.86 | 0.64 | 0.79 | 0.60 | 0.89 |